\documentclass[a4paper,fleqn,usenatbib]{mnras}

\usepackage[T1]{fontenc}
\usepackage{ae,aecompl}

\usepackage{verbatim,graphicx,epsfig,dcolumn,color,hyperref}

\usepackage{graphicx}	
\usepackage{amsmath}	
\usepackage{amssymb}	

\newcolumntype{.}{D{.}{.}{4}}
\newcolumntype{,}{D{.}{.}{2}}
\newcolumntype{;}{D{.}{.}{1}}
\newcommand{\nodata}{$\cdot\cdot\cdot$}

\title[Pre-discovery \emph{Hipparcos} transits of exoplanets]{Pre-discovery transits of the exoplanets WASP-18 b and WASP-33 b from \emph{Hipparcos}}
\author[I. McDonald et al.]{I.~McDonald$^{1}$\thanks{E-mail: mcdonald@jb.man.ac.uk}, E.~Kerins$^{1}$\\
$^{1}$Jodrell Bank Centre for Astrophysics, Alan Turing Building, Manchester, M13 9PL, UK\\
}

\date{Accepted XXX. Received YYY; in original form ZZZ}

\pubyear{2018}

\begin{document}
\label{firstpage}
\pagerange{\pageref{firstpage}--\pageref{lastpage}}
\maketitle

\begin{abstract}
We recover transits of WASP-18\,b and WASP-33\,b from \emph{Hipparcos} (1989--1993) photometry. Marginal detections of HAT-P-56\,b and HAT-P-2\,b may be also present in the data. New ephemerides are fitted to WASP-18\,b and WASP-33\,b. A tentative ($\sim$1.3$\sigma$) orbital decay is measured for WASP-18\,b, but the implied tidal quality factor ($Q^\prime \sim 5 \times 10^5$) is small and survival time ($< 10^6$ years) is too short to be likely. No orbital decay is measured for WASP-33\,b, and a limit of $Q^\prime > 2 \times 10^5$ is placed. For both planets, the uncertainties in published ephemerides appear underestimated: the uncertainty in the period derivative of WASP-18\,b would be greatly reduced if its current ephemeris could be better determined.
\end{abstract}



\section{Introduction}
\label{IntroSect}

Exoplanetary science is a relatively young field, hence many long-term evolutionary characteristics of planetary systems remain unknown. Pre-discovery archival data can provide, e.g., more precise orbital properties. Changes in these properties may come from transit timing variations (TTVs) caused by a second planet in the system \citep[e.g.][]{SFA+13}, or by long-term orbital expansion or decay, due to stellar mass loss or tidal inspiral \citep[e.g][]{MV12}. In particular, historical data lets us constrain the tidal quality factor of exoplanet hosts, allowing us to model tidal effects from stars more generally \citep[e.g.][]{PJST12}.

Few historical observations have sufficient sensitivity or cadence to detect exoplanets. Photometric accuracy of better than $\sim$0.01 mag is generally required, while duty cycles of transits are typically only a few per cent of the orbit, so dozens of repeated visits are necessary to secure a transit. Of the literature data available, only the \emph{Hipparcos} satellite \citep{Perryman97,vanLeeuwen07} has sufficient accuracy and cadence to reliably search for exoplanets \emph{en masse}. \emph{Hipparcos} operated between 1989 and 1993, and returned broadband photometry to an accuracy of a few millimagnitudes on around 120\,000 nearby stars. Transits of HD\,209458\,b and HD\,189733\,b have had their \emph{Hipparcos} photometry published already \citep{RA00,HLDE06}. In this article, we search for transits of other known exoplanets in the original \emph{Hipparcos} data\footnote{VizieR catalogue I/311}.


\section{Transiting exoplanets in the \emph{Hipparcos} dataset}
\label{PlanSect}

\begin{center}
\begin{table*}
\caption{\emph{Hipparcos} stars exhibiting transits of $>$5 mmag.}
\label{HipTable}
\begin{tabular}{@{}lrclrrrrr@{}c@{}r@{}r@{}}
    \hline \hline
\multicolumn{1}{c}{Name} & \multicolumn{1}{c}{HIP} & \multicolumn{1}{c}{$d$} & \multicolumn{1}{c}{$T_0$} & \multicolumn{1}{c}{$P$} & \multicolumn{1}{c}{$T_{14}$} & \multicolumn{1}{c}{Depth} & \multicolumn{1}{c}{\emph{Hip.} rms}  & \multicolumn{1}{c}{$N$}  & \multicolumn{1}{c}{Expected}   & \multicolumn{2}{c}{Observed} \\
          \    & \     & \multicolumn{1}{c}{\ } & \multicolumn{1}{c}{(TJD)} & \multicolumn{1}{c}{\ }  & \multicolumn{1}{c}{\ }  & \multicolumn{1}{c}{\ } & \multicolumn{1}{c}{\ }   & \multicolumn{1}{c}{\ }   & \multicolumn{1}{c}{detection}   & \multicolumn{2}{c}{depth} \\
          \    & \     & \multicolumn{1}{c}{(pc)} & \multicolumn{1}{c}{(d)} & \multicolumn{1}{c}{(d)}  & \multicolumn{1}{c}{(d)}  & \multicolumn{1}{c}{(mmag)} & \multicolumn{1}{c}{(mmag)}   & \multicolumn{1}{c}{\ }   & \multicolumn{1}{c}{($\sigma$)}   & \multicolumn{1}{c}{($\sigma$)}	& \multicolumn{1}{c}{\llap{(}mmag\rlap{)}} \\
    \hline
        WASP-18\,b &	7562  & 126 $\pm$ 5	& 4644.90531 & 0.94145299 & 0.09089 & 9.16  & 18.8 &  9 & 1.38 &   2.81 &  14.6 \\
	WASP-33\,b & 	11397 & 118 $\pm$ 3	& 4163.22373 & 1.21986975 & 0.11224 & 11.36 & 12.9 & 15 & 3.29 &   3.12 &  10.8 \\
	HD\,17156 b & 	13192 & 79.8 $\pm$ 1.6	& 4756.7313  & 21.21663   & 0.1338  & 5.29  & 19.0 &  0 & 0.00 & \nodata&\nodata\\ 
	KELT-7\,b & 	24323 & 138 $\pm$ 5	& 6223.9592  & 2.7347749  & 0.14630 & 8.28  & 16.7 &  0 & 0.00 & \nodata&\nodata\\ 
	KELT-2\,A\,b & 	29301 & 134 $\pm$ 6	& 5974.60335 & 4.113791   & 0.2155  & 5.21  & 21.0 &  2 & 0.25 & --0.79 &--16.6 \\
	HAT-P-56\,b & 	32209 & 319 $\pm$ 23	& 6553.61645 & 2.7908327  & 0.09463 & 11.11 & 15.1 &  6 & 0.74 &   0.83 &  12.5 \\
	HD 80606\,b & 	45982 & 65.2 $\pm$ 1.1	& 4876.344   & 111.43670  & 0.504   & 11.17 & 16.7 &  0 & 0.00 & \nodata&\nodata\\ 
	GJ\,436\,b & 	57087 & 10.1 $\pm$ 0.2	& 4415.62074 & 2.643850   & 0.03170 & 6.96  & 75.0 &  0 & 0.00 & \nodata&\nodata\\ 
	HAT-P-2\,b & 	80076 & 129 $\pm$ 4	& 4397.49375 & 5.6334729  & 0.1787  & 5.22  & 19.7 &  9 & 0.75 &   0.94 &   6.6 \\
	HD\,189733\,b & 98505 & 19.8 $\pm$ 0.1	& 4279.43671 & 2.21857567 & 0.0760  & 24.12 & 15.1 &  4 & 2.76 &   3.73 &  32.6 \\
	HD\,209458\,b & 108859&	48.9 $\pm$ 0.5	& 2826.62851 & 3.52474859 & 0.1277  & 14.61 & 14.8 &  5 & 1.97 &   3.56 &  26.4 \\
    \hline
\multicolumn{12}{p{0.95\textwidth}}{Notes: Distances come from \emph{Gaia} Data Release 1 \citep{GaiaDR1}, with the exception of GJ 436, which comes from \citet{vanLeeuwen07}. Transit parameters are sourced from the EDE (values for WASP-18\,b and WASP-33\,b explicitly come from \citet{WDB+17} and \citet{ZKK+17}); truncated Julian dates are given as TJD = JD $-$ 2\,450\,000 days. $N$ is the number of observations expected during transit.}\\
    \hline
\end{tabular}
\end{table*}
\end{center}

Exoplanets in the \emph{Hipparcos} dataset were selected from the Exoplanets Data Explorer (EDE\footnote{\url{http://exoplanets.org}}; \citealt{HWW+14}), using the parameters ``{\tt TRANSIT == 1 \&\& HIPP > 0}''. This returned 17 unique systems. We further restricted our criteria to a transit depth $>$5 mmag ({\tt DEPTH > 0.005}), returning the 11 systems listed in Table \ref{HipTable}.

For HAT-P-56 and HD 189733, outliers in the \emph{Hipparcos} data were removed using a $\kappa\sigma$-clipping routine: i.e., an iterative pass of the data was performed, removing points more than $\kappa$ standard deviations from the mean. A cutoff of $\kappa = 3.5$ was applied, which was chosen so as not to remove points in the expected transit regions. As stars have between 54 and 187 data points, any choice of $\kappa \gtrsim 2.7$ is not expected to remove valid data from the fit.

The photometric data were folded on literature orbit ephemerides (Table 1). Four transiting planets were expected to be detected ($>$1$\sigma$): WASP-18\,b, WASP-33\,b, HD\,189733\,b and HD\,209458\,b and all four were recovered. Transits of KELT-2\,A\,b were not recovered due to the low signal-to-noise ratio. Transits of HAT-P-56\,b and HAT-P-2\,b were expected just below the 1$\sigma$ detection limit, and measurements of the recovered transit depth are close to the 1$\sigma$ limit. Since this measurement effectively uses a boxcar transit, and since the \emph{Hipparcos} photometric transmission curve is relatively blue ($\lambda_{\rm eff} \approx 5275$ \AA), a limb-darkened model is expected to recover these transits at just above 1$\sigma$. However, since the photometry would be of insufficient quality to model further, they are neglected for the remainder of this paper.

\begin{figure}
\centerline{\includegraphics[height=0.45\textwidth,angle=-90]{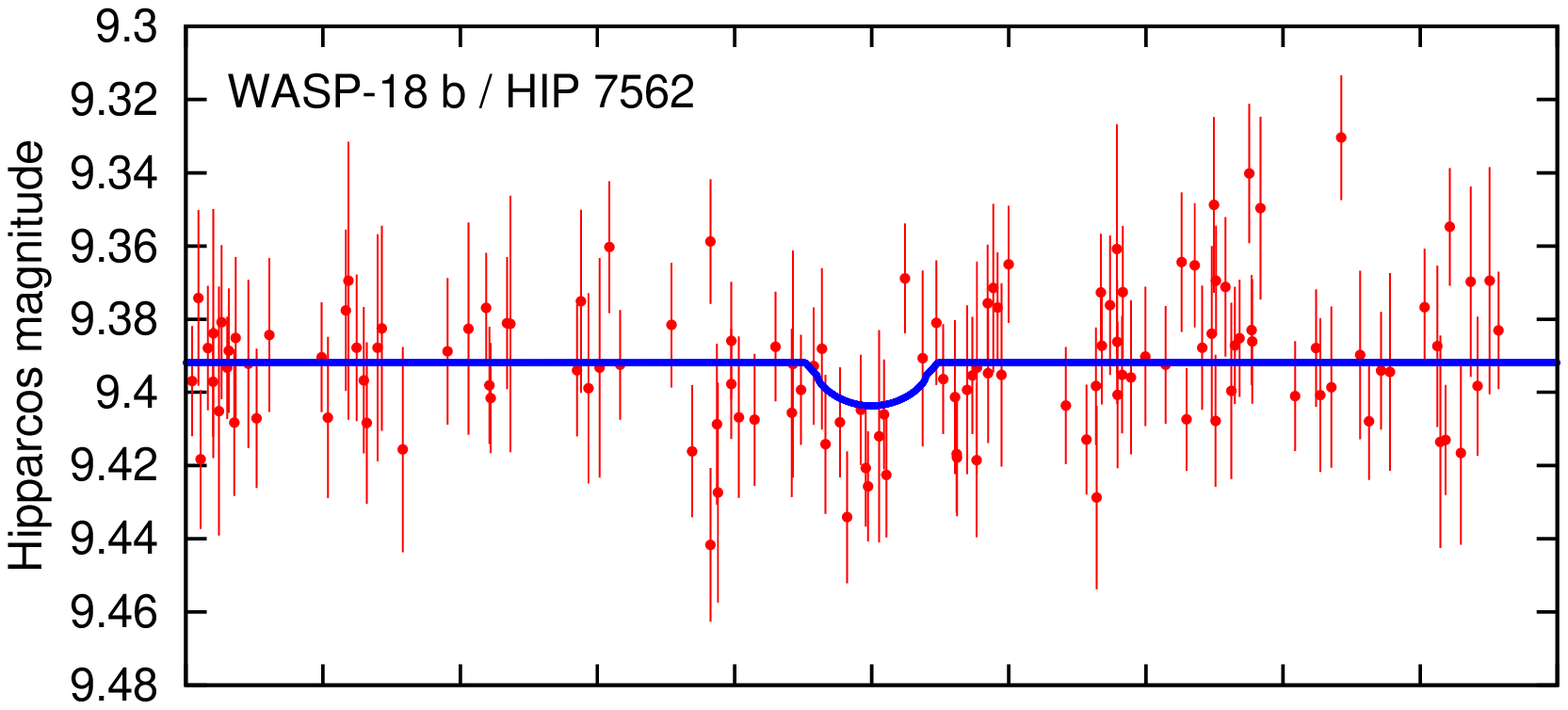}}
\centerline{\includegraphics[height=0.45\textwidth,angle=-90]{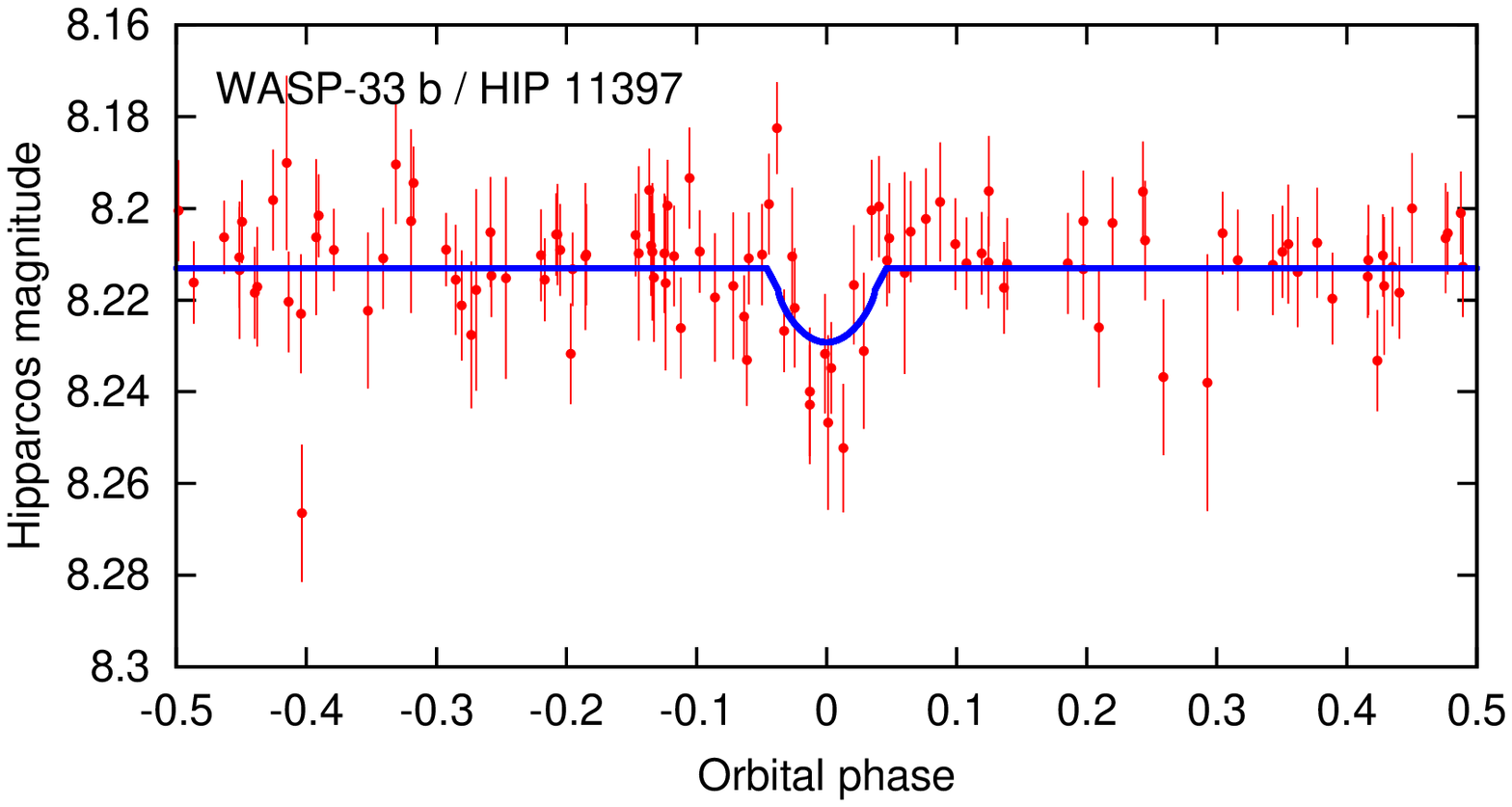}}
\caption{\emph{Hipparcos} photometry, phase-folded on a modern ephemeris. Lines show the expected transit position, width and depth.}
\label{HipFig}
\end{figure}

WASP-18\,b and WASP-33\,b have never previously been recovered from \emph{Hipparcos} data. Their lightcurves are shown in Figure \ref{HipFig}, folded on the empherides from Table \ref{HipTable}. Data sampling is sparse: 132 points over 1190 days for WASP-18\,b and 113 points over 930 days for WASP-33\,b (one point has been cleaned by $\kappa\sigma$-clipping from the latter). Consequently, a blind search for planets in the \emph{Hipparcos} data would have been liable to miss these transits, which are not apparent in the unfolded lightcuves.

\begin{figure}
\centerline{\includegraphics[height=0.45\textwidth,angle=-90]{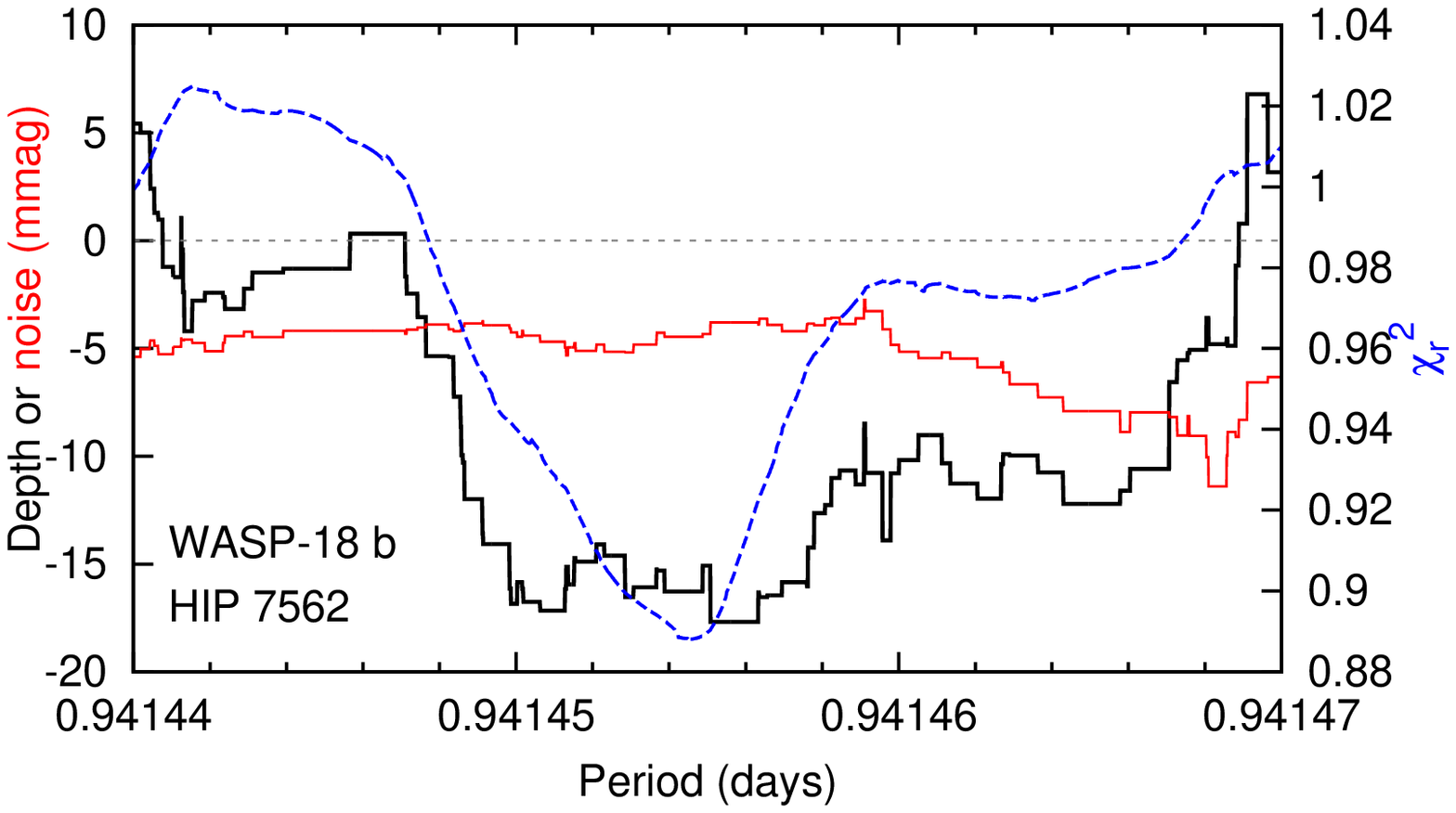}}
\centerline{\includegraphics[height=0.45\textwidth,angle=-90]{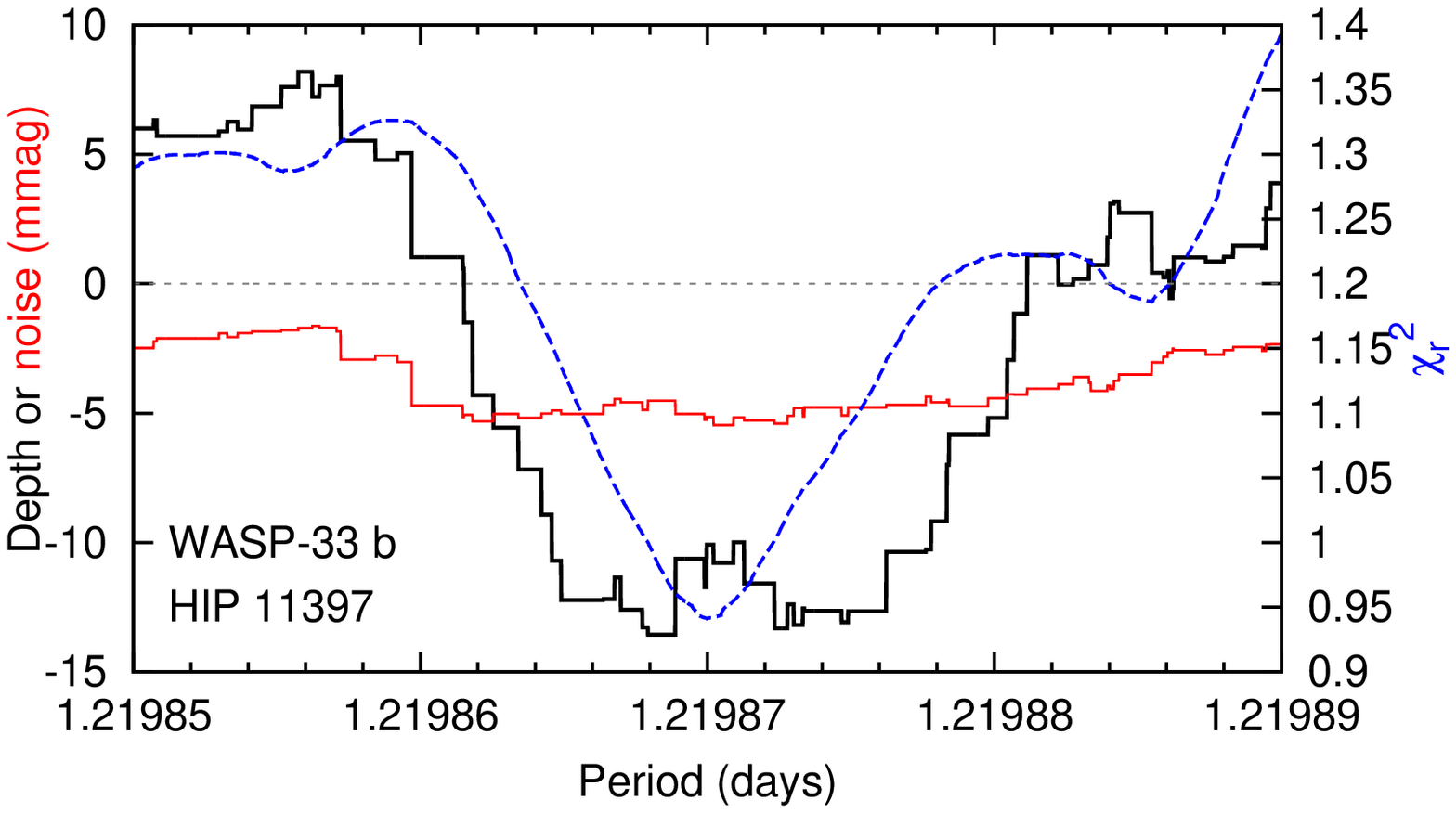}}
\caption{Goodness-of-fit of \emph{Hipparcos}-derived periods. Dark, black lines show the mean transit depth (across $t_{14}$) at that period; thin, red lines show the out-of-transit noise level. Dashed, blue lines show the reduced $\chi^2$ using the limb-darkened model (right axis).}
\label{FitFig}
\end{figure}

\begin{figure}
\centerline{\includegraphics[height=0.45\textwidth,angle=-90]{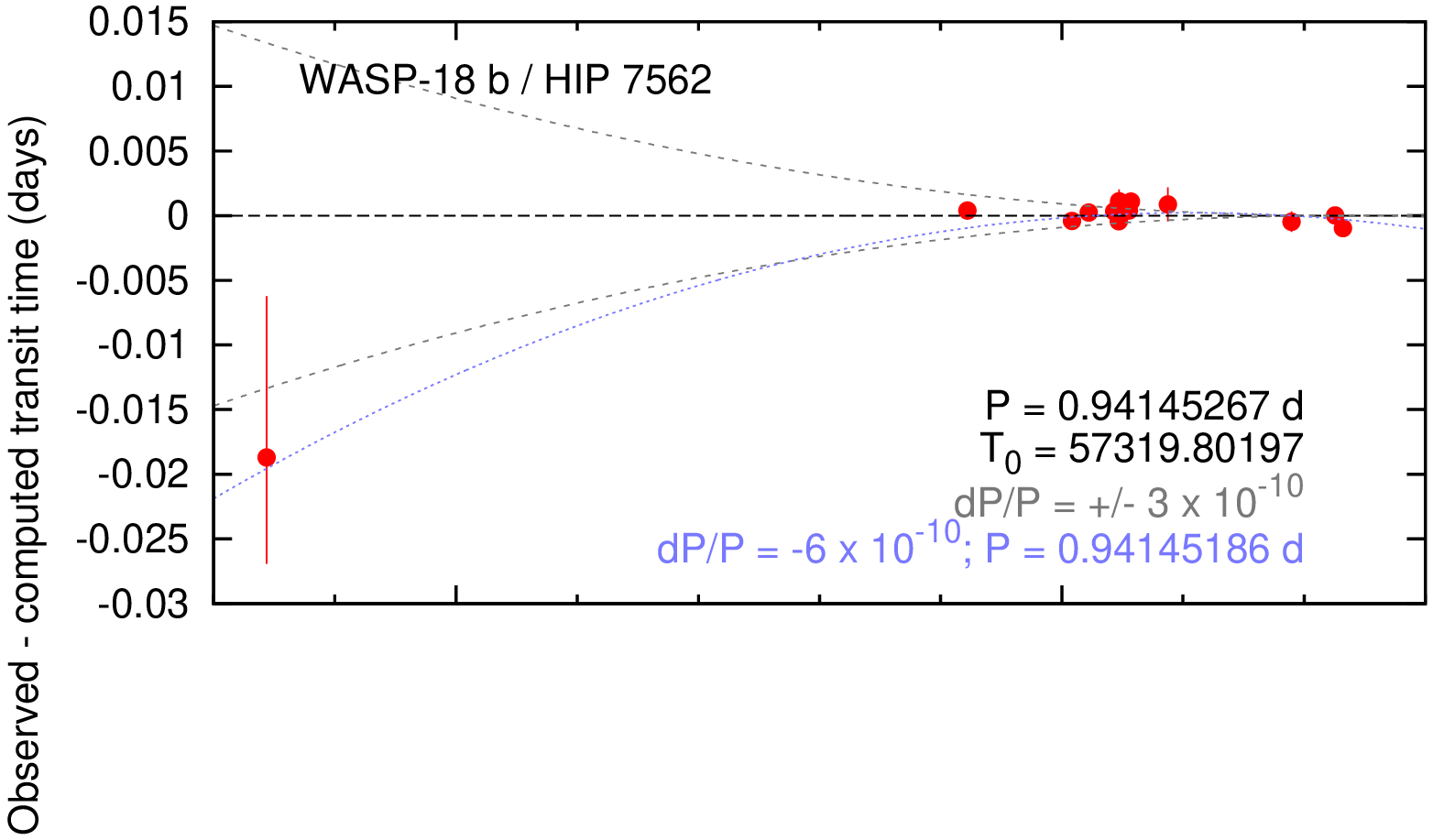}}
\centerline{\includegraphics[height=0.45\textwidth,angle=-90]{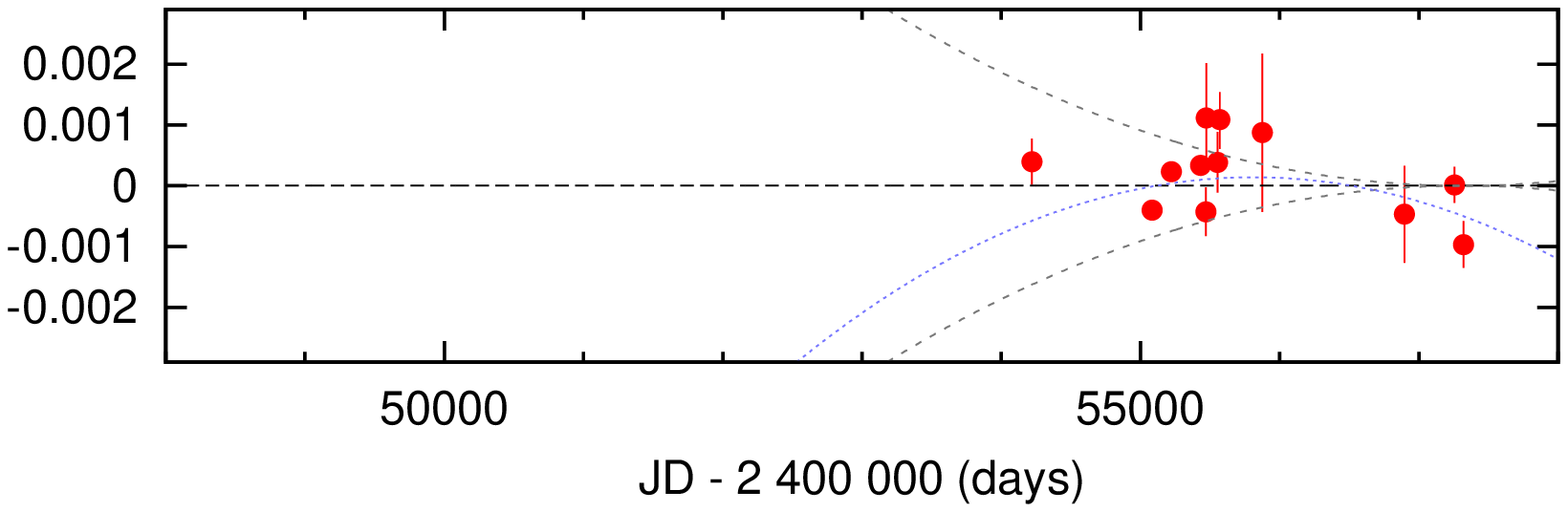}}
\centerline{\includegraphics[height=0.45\textwidth,angle=-90]{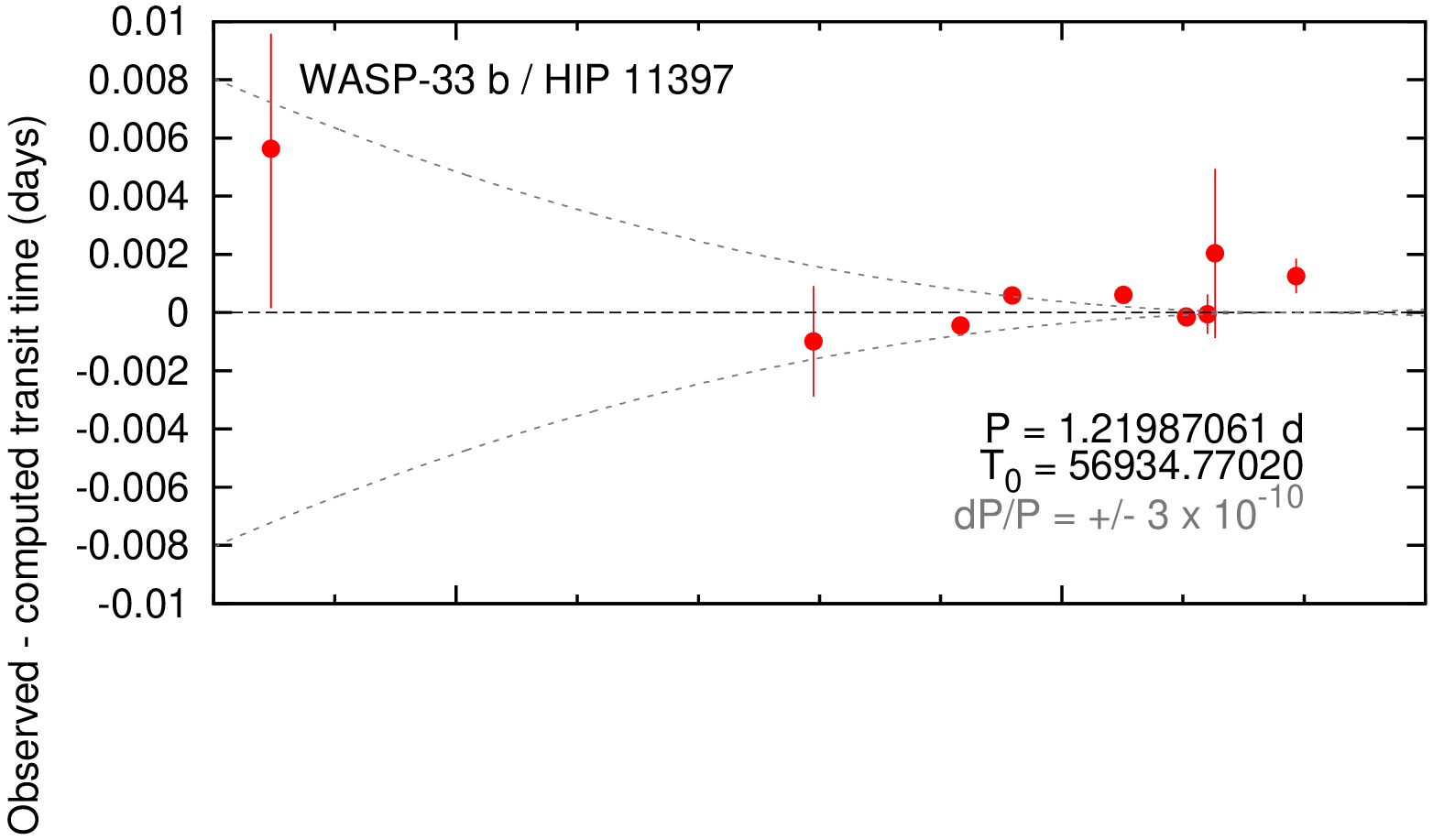}}
\centerline{\includegraphics[height=0.45\textwidth,angle=-90]{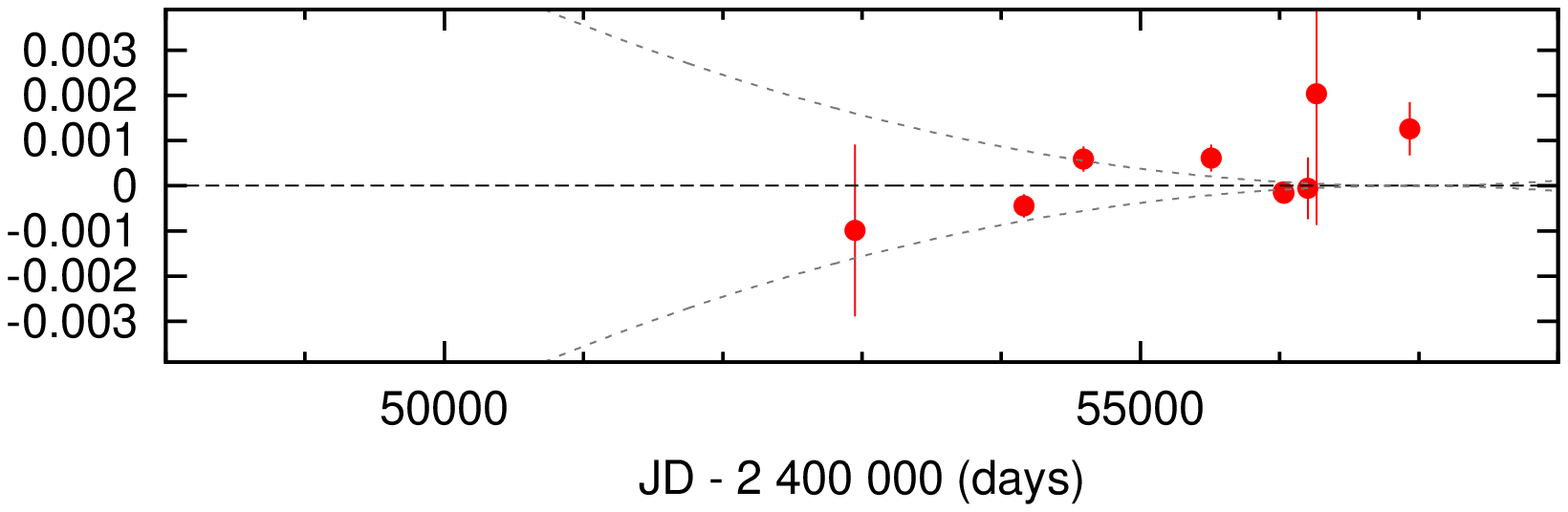}}
\caption{$O-C$ diagrams for WASP-18 b and WASP-33 b, modelled against the best-fit ephemeris. Curves show models with period changes (d$P/P$), as indicated on each plot.}
\label{OMCFig}
\end{figure}


\section{Orbital solutions and evolution}
\label{OrbitSect}




As in previous analyses of \emph{Hipparcos} photometry \citep{RA00,HLDE06}, we note that fitting a two-parameter ephemeris (mid-transit epoch and period, $T_0$ and $P$) to data of this quality is less accurate than taking an established ephemeris and providing a refined period. In each case, $T_0$, $t_{14}$ and $R_{\rm p}/R_\ast$ were held fixed to the values in Table \ref{HipTable}, and the \emph{Hipparcos} data were folded on a range of periods spanning 0.000015 days either side of these ephemerides.

The transit was represented by a trapezoid ingress and egress, based on the above parameters. The impact of including limb darkening on the precision of the resulting fit was found to be significant, but the exact treatment of limb darkening was not. Hence, the transit between second and third contact was modelled as a point source crossing a limb-darkened star, with limb-darkening co-efficients taken from {\sc jktld} \citep{Southworth08}: inputs of $T_{\rm eff} = 6400$ and 7430 K, $\log(g) = 4.367$ and 4.300 dex and [Fe/H] = 0.0 and 0.1 dex were assumed for WASP-18 and WASP-33, respectively, while a microturbulent velocity of 2 km s$^{-1}$ and a quadratic law with \citet{Claret04} models was assumed for both, and the \emph{Hipparcos} filter was approximated by Sloan $g^\prime$. A $\chi^2$ minimisation performed to identify allowed periods for the \emph{Hipparcos} data. The reduced $\chi^2$ minimum is close to unity in both cases (Figure \ref{FitFig}), so the periods where $\chi^2 \leq \chi^2_{\rm min}+1$ can be used to approximate the period uncertainty. The differences between light curves with transiting planets and flat light curves are $\Delta\chi^2 = 12$ and 16 for WASP-18\,b and WASP-33\,b, respectively, so the transits are detected with clear significance. The fitted periods and corresponding mid-transit times for this two-epoch fit are
\begin{itemize}
\item $P = 0.941\,454\,55 \,^{+0.000\,000\,87} _{-0.000\,001\,32}$ days, and
\item $T_0 = 2\,448\,436.2359 \,^{+0.0125} _{-0.0082}$
\end{itemize}
for WASP-18\,b and
\begin{itemize}
\item $P = 1.219\,869\,98 \,^{+0.000\,000\,79} _{-0.000\,000\,57}$ days, and
\item $T_0 = 2\,448\,472.5334 \,^{+0.0040} _{-0.0055}$
\end{itemize}
for WASP-33\,b.

These mid-transit times represent observations taken 16 years (6145 and 4665 orbits) before those in the discovery papers of each planet \citep{HAC+09,CCGS+10}, and more than double the length of their observational record to 24 and 23 years, respectively. To these transit times, we added the literature transit photometry collated for both WASP-18\,b and WASP-33\,b (\citealt{WDB+17} and \citet{ZKK+17}, respectively), and created $O-C$ diagrams for each planet (Figure \ref{OMCFig}).

Unfortunately, the low cadence of the \emph{Hipparcos} compared to modern data means that they do not provide constraints greatly better than those available in the current literature \citep{TPB+16,ZKK+17,WDB+17}.

To fit the orbits, we ran two-parameter ($T_0, P$) and three-parameter ($T_0, P, \delta P / P$) Monte-Carlo $\chi^2$ fits to the observed mid-transit times. A two-parameter fit for this entire dataset formally provides
\begin{itemize}
\item $P = 0.941\,452\,67 \pm 0.000\,000\,11$ days,
\item $T_0 = 2\,457\,319.80197 \pm 0.00021$, and
\item $\chi^2_r = 5.14$
\end{itemize}
for WASP-18\,b and
\begin{itemize}
\item $P = 1.219\,870\,61 \pm 0.000\,000\,15$ days,
\item $T_0 = 2\,456\,934.77020 \pm 0.00010$, and
\item $\chi^2_r = 2.50$
\end{itemize}
for WASP-33\,b. These fits are shown in the $O-C$ diagrams in Figure \ref{OMCFig}. A three-parameter fit formally provides:
\begin{itemize}
\item $\delta P/P = -6 \pm 2 \times 10^{-10}$,
\item $P = 0.941\,451\,86 \pm 0.000\,000\,23$ days,
\item $T_0 = 2\,457\,319.80167 \pm 0.00026$, and
\item $\chi^2_r = 4.64$
\end{itemize}
for WASP-18\,b and
\begin{itemize}
\item $\delta P/P = 2 \pm 3 \times 10^{-10}$,
\item $P = 1.219\,870\,93 \pm 0.000\,000\,50$ days,
\item $T_0 = 2\,456\,934.77090 \pm 0.00017$, and
\item $\chi^2_r = 2.78$
\end{itemize}
for WASP-33\,b. The fit for WASP-18\,b is shown as the dotted line in Figure \ref{OMCFig}.


\section{Discussion and conclusions}
\label{DiscSect}

The reduced $\chi^2$ minimum of these fits is substantially greater than unity: in both bodies, an unmodelled scatter of around 0.001 days (1.44 minutes) is seen in the $O-C$ diagrams. This suggests that the errors quoted above are likely to be underestimates, either due to physical or unmodelled instrumental sources (cf.\ \citealt{ALME+10,BBG+13}). It also implies that either the photometric uncertainties on the input data are underestimated, or that an undetected third body in the system is causing TTVs.

We used {\sc ttvfaster} \citep{AD16} to model a third-body TTV signal to the data, assuming a circular orbit, co-planar to the relevant planet. Unfortunately, the only ranges of parameters that can produce a sufficiently strong signal ($\Delta T_{\rm TTV} \gtrsim 0.0005$ days) are of dynamically unstable systems, or those where a companion would be spectroscopically detectable (e.g.\ a 0.25 M$_\odot$ star in a 5.7-day orbit). Unless cyclical variation of the planets' orbits are being driven by tidal interaction with their host stars, it appears that the uncertainties on the published transit times have been under-estimated in several cases, which could be due in part to microvariability on the host stars \citep{vECW+14}.

Given these under-estimated uncertainties, and the possibility of other physical sources of TTV, the significance of the orbital change of either exoplanet cannot be precisely computed. Taking only the \emph{Hipparcos} data at face value, we have a $\sim$1.0$\sigma$ measurement of orbital expansion in WASP-33 b, and a $\sim$1.3$\sigma$ measurement of orbital decay in WASP-18 b, depending on the exact period adopted. These are not significant detections.

Strong orbital decay is not expected for these planets, as their host stars are relatively warm and have thin convective envelopes in which tides can be generated. The tidal quality factors for these stars are expected to be $Q^\prime \sim 10^8$ \citep{BO09}. Due to its spin-orbit misalignment \citep{CCGS+10}, non-radial changes to the orbit of WASP-33 may also be expected \citep{Iorio11,LO17}, causing more complex TTV signals over long periods.

Using Equations 4 and 5 of \citep{WDB+17}, $\delta P/P \sim -6 \times 10^{-10}$ implies $Q^\prime \sim 5 \times 10^5$ for WASP-18. This is a much smaller value than nominally expected (cf.\ \citealt{CCJ18,PBWH18}), but similar to that proposed for WASP-12\,b by \citet{MDF+16}. However, it also implies a survival time of $<10^6$ years, thus the mere observable presence of WASP-18\,b means this value of $\delta P/P$ is likely to be erroneously high. For WASP-33, $\delta P/P < -1 \times 10^{-10}$ implies $Q^\prime > 2 \times 10^5$, which is not very limiting, but interesting given the visible tides the planet generates on its star \citep{vECW+14}.

A significant uncertainty driving the difference between the two- and three-parameter fits for WASP-18\,b is the period in the current epoch, which differs by $\sim 8 \times 10^{-7}$ days. A few high-precision measurements of transit times in the current epoch could greatly constrain these uncertainties, determining whether the offset of the \emph{Hipparcos} datapoint in the $O-C$ diagram is significant. We therefore strongly encourage monitoring of WASP-18 b, to more accurately determine its orbital period in the present epoch.

\section*{Acknowledgements}

The authors acknowledge support from the UK Science and Technology Facility Council under grant ST/P000649/1.



\bsp	
\label{lastpage}
\end{document}